\documentclass[12pt]{article}
\usepackage{epsf}
\usepackage{slashed}
\usepackage{color}
\begin{document}
\begin{titlepage}
\title{Quarks, Hadrons, and Emergent Spacetime}
\author{
{Piotr \.Zenczykowski }\footnote{E-mail: piotr.zenczykowski@ifj.edu.pl}\\
{\em Division of Theoretical Physics}\\
{\em the Henryk Niewodnicza\'nski Institute of Nuclear Physics}\\
{\em Polish Academy of Sciences}\\
{\em Radzikowskiego 152,
31-342 Krak\'ow, Poland}\\
}
\maketitle
\begin{abstract}
It is argued that important information on the emergence of space is hidden at the quark/hadron level. The arguments follow from the acceptance of the conception that space is an attribute of matter. They involve in particular the discussion of possibly relevant mass and distance scales,  
the generalization of the concept of mass as suggested by the phase-space-based  explanation of the rishon model, and the phenomenological conclusions on the structure of excited baryons that are implied by baryon spectroscopy. A counterpart of the Eddington-Weinberg relation concerning Regge towers of hadronic resonances is noted.

\end{abstract}

\vfill
\end{titlepage}

\section{Introduction}
\label{sec1}
{Modern fundamental physics is dominated by two divergent lines of philosophical thought concerning the notion of space. Both these lines originated in ancient Greece. They have led us to two disjoint but very successful theories that describe the behaviour of matter `in the small' and `in the large', the Standard Model (SM) and General Relativity (GR). The SM, a relativistic quantum field theory of interacting elemantary particles, is a descendant of the Democritean line of thought which accepts separation of matter and space, and views space as a mere container in which indivisible `atoms' move. 
General Relativity is a remote heir of Aristotle's way of thinking in which space
(or rather `place') is regarded as an attribute of matter. Such a logical priority of matter over space is a cornerstone of the views of many modern philosophically-minded thinkers such as Leibniz or Mach and contributed to Einstein's development of GR.}
\\

Now, if one accepts the latter position, i.e. the philosophically very attractive view that `space and time are (...) stretched out by matter' \cite{Heisenberg1979}, that `objects make space', that without matter there is no space, properties of space should follow those of matter. Accordingly, quantum properties of matter and
the discretization of mass prompt us to expect some form of discretization (or quantization) of space. 
Such arguments and the  general wish to unite the classical and quantum aspects of reality lead to the idea of quantum gravity, according to which the macroscopic classical continuous space is  to be replaced in the microcosm by some form of quantized space. A simple dimensional argument which involves the gravitational, 
quantum and relativistic constants $G = 6.67 \times 10^{-8}~cm^3/(g~s^2)$, $h = 6.62 \times 10^{-27}~g~cm^2/s$, and $c = 3\times 10^{10}~cm/s$, and singles out the Planck units of
length $l_P =\sqrt{hG/c^3} = 4.05 \times 10^{-33}~cm$ and time $t_P= l_P/c$ (as well as  mass $m_P = \sqrt{hc/G} = 5.46 \times 10^{-5}~g$), is then widely accepted 
as providing the distance and time scales at which the quantum nature of space should manifest. \\

One of the problems associated with the idea of the emergence of macroscopic continuous space at Planck's length is its diminutive size and the resulting lack of.
experimental information that
 could guide our theoretical speculations.  Instead, the available experimental input is located in 
the particle  sector, at distance scales that are generally deemed irrelevant for the idea. Indeed, with field-theoretical approaches describing the behaviour of elementary particles fairly well, the idea to link the particle sector with the concept of the emergence of space may appear far-fetched.
Yet - although something may indeed happen at the Planck distance scale -  important information 
on the idea of space emergence may still reside
at the particle and hadronic scales.
After all, with space regarded as an attribute (or derivative) of matter, and with the discretization of mass hinting at some form of the discretization of space, it is the
variety of particles and the spectrum of their masses that should
direct our ideas on the quantization of space.

\section{Mass and distance scales}
\label{sec2}

\subsection{Planck scale}

The lack of experimental input that could direct theoretical research on quantum gravity raises various questions. In particular, one may doubt both the need for the existence of the underlying quantized space and the relevance of Planck scale to such ideas.
Indeed, gravity may be a strictly classical long-distance (residual?) phenomenon that does not have an underlying quantum counterpart \cite{Wuthrich}. Furthermore, as observed by some authors \cite{Amelino} and as clearly voiced by Meschini \cite{Meschini}, the only links between the general idea of an underlying quantized space 
and the Planck natural units are the dimensional analysis and the theoretical ad hoc considerations of Planck-size black holes. However, as pointed out in \cite{Meschini}, dimensional analysis is not a 
trustworthy tool that could provide us with reliable information on the realm of the unknown.  This claim is justified in \cite{Meschini} with a few examples from the history of physics which show 
that dimensional analysis becomes reliable only when the underlying theory is already known.  For example, Meschini considers estimates of the size of hydrogen's atom and its binding energy that could have been made on the basis of dimensional analysis before any knowledge of Planck constant and the associated quantum features of reality were available. 
Assuming that hydrogen atom is built from electron and proton and accepting that the relevant dimensional constants are those involved in the classical theories of mechanics and electromagnetism, he shows that the predicted numbers are some five orders of magnitude away from their physical values. \\

On the other hand, the field-theoretical approach of the Standard Model works quite well down to the distances of the order of some $10^{-16}$~cm, three
orders of magnitude below typical hadronic size. On this basis one may hope that no change in the properties of space should occur for still smaller distances and that this trend continues down to the naturally distinguished scale of Planck's length. 
Yet, such a point of view ignores the fact that our conclusions concerning properties of space at small distances do not follow from direct observations but are inferred with the help of quantum field theory (which itself has a hybrid classical-quantum nature
\cite{Finkelstein}).   In particular, it also ignores the issue of quantum nonlocality which questions the classical conception of an underlying spacetime even at macroscopic distances. Of course, quantum nonlocality does not directly jeopardize relativistic local quantum field theory.  
Yet, there is a clear tension between the classical and quantum descriptions, between locality and nonlocality. The deep and unaswered question is how classical locality emerges out of quantum nonlocality.\\

Now, experiments show that nonlocal quantum correlations appear
at all probed distance scales. It looks as if spacial distances were completely irrelevant at the `true' quantum level.
It seems that the macroscopic spacetime of the classical description of nature emerges from the underlying quantum layer in a subtle and intrinsically nonlocal (holographic?) way, 
as the recent idea of building spacetime from quantum entanglement also suggests.
There does not seem to be a reason to choose the Planck distance scale over the particle/hadronic scale (or the universe scale)
 as more appropriate for the emergence of the classical spacetime. From the vantage point of quantum nonlocality the Planck distance scale seems to lose its priviliged status. 
In fact,  it was argued in the quantum gravity context \cite{Bojowald} that the typical distance scale $l_{QS}$ relevant for the 
idea of space quantization may be dynamical and much larger than $l_P$:  $l_P << l_{QS} << l_U=c H^{-1}_0$ (where $l_U$ is the observable radius of the Universe and $H_0$ is the Hubble constant), and that the corresponding mass scale 
$m_{QS} = h/(c~l_{QS})$ is much smaller than $m_P$.

\subsection{Hadronic scale}
 The case for space quantization and emergence may look altogether different when other fundamental constants are admitted into the game.
For example, there seems to be no reason why 
cosmological constant $\Lambda = 1.19 \times 10^{-56}~cm^{-2}$ ($\approx (1/l_U)^2$)
should not be regarded as a constant as important as $h$, $G$, and $c$. If $\Lambda$ is relevant, one can form 
(after Wesson \cite{Wesson}) two additional mass scales that differ from $m_P$, namely
\begin{equation}
\label{mW}
m_W = (h/c) \sqrt{\Lambda/3} = 1.39 \times 10^{-65}~g,
\end{equation}
and
\begin{equation}
\label{mU}
m_U = (c^2/G) \sqrt{3/\Lambda} = 2.14 \times 10^{56}~g,
\end{equation}
with the Planck mass being their geometrical mean
\begin{equation}
\label{mP2}
m^2_P = m_W m_U.
\end{equation}
Masses $m_U$ and $m_W$ are connected by an enormous dimensionless quantity:
\begin{equation}
\label{N10to120}
N = m_U/m_W = 3 c^3/(hG\Lambda) = 15.4 \times 10^{120}=m^2_U/m^2_P.
\end{equation}
The Wesson mass $m_W$, being proportional to $h$ and independent of $G$, may be 
interpreted as the quantum of mass. The other mass, $m_U$, which is independent of $h$,
and thus seems appropriate for the classical limit,
is of the order of the mass of the observable Universe. Now,
with Planck mass being around one tenth of the mass of a flea, one may convincingly argue
that $m_P$ corresponds to the classical and not the quantum realm. What meaning should be 
assigned then to Planck length which is directly related to $m_P$ by standard
quantum considerations: $l_P = h /(m_P c)$? Does it make sense to apply quantum ideas 
to flea-weight objects in this way? Why is the interest of the majority
focused on space and tiny distances ($l_P$), and not on matter and tiny masses (e.g. $m_W$)?
Shouldn't the length and mass scales $l_{QS}$ and $m_{QS}$ be both far 
from the classical realm?
\\

It may be observed that mass scales $m_W$, $m_P$, and $m_U$ are defined by expressions in which one of the four fundamental constants is missing (respectively: $G$, $\Lambda$, and $h$) \cite{Thai}. Thus, one
additional mass scale (the one without $c$) may be singled out as potentially important. It is
\begin{equation}
\label{N1/3}
m_N = m_W N^{1/3} = m_U N^{-2/3}
= \left( (h^2/G) \sqrt{\Lambda/3}\right) ^{1/3} 
= 0.346 \times 10^{-24}~{g}. 
\end{equation}
This number, which is smaller than the Planck mass by some 20 orders of magnitude, i.e.
\begin{equation}
m_P=m_NN^{1/6},
\end{equation}  
should be compared with the electron, pion and nucleon masses:
$m_e = 0.91 \times 10^{-27} ~g$, $m_{\pi} = 0.25 \times 10^{-24}~g$,
$m_n = 1.67 \times 10^{-24} g$.
The fact that $m_N$, the particular combination of $h$, $G$, and $\Lambda$, is of the order of $m_n$ (or $m_{\pi}$) was noticed
by many authors and is sometimes called the Eddington-Weinberg relation \cite{EW} (see also \cite{Thai,Funkhouser}).  
Furthermore, the distance scale related to $m_N$ by typical quantum
considerations, ie.
$l_N = h/(m_N c) = 6.37 \times 10^{-13}~{\rm cm}$,
is obviously of the order of typical hadronic size. 
Note that, unlike $l_P$ and $m_P$, {\it both} the distance and mass scales $l_N$ and $m_N$ are  relatively small with 
respect to typical classical macroscopic distances and masses.
Are hadronic scales relevant for the idea of space quantization? \\

In fact, there is not one but two a priori independent mass-related parameters that describe the hadronic spectrum. The first is the mass of the lowest lying hadronic states such as nucleon that fits the Eddington-Weinberg relation.
The other parameter is concerned with the pattern of excited hadronic states. Here, the relevant feature of the hadronic spectrum is the appearance of linear Regge trajectories which describe infinite `towers' of similar hadronic
resonances (i.e. the recurrences of the ground state mesons and baryons with increasing spins $J$ and masses $m$) according to the generic formula:
\begin{equation}
\label{Regge}
J = \alpha_0 + \alpha ' m^2, 
\end{equation}
where $\alpha_0$ is the `intercept' and $\alpha'$ - the slope of the trajectory.
The slopes of all trajectories are similar: $\alpha' \approx 0.9~{GeV}^{-2}$ (with masses measured in $GeV$'s, the units of energy).
For example, for nucleon one has $1/2 = \alpha_{0,n} + \alpha ' m_n^2$
(i.e. the intercept of nucleon trajectory is
  $\alpha_{0,n} \approx -1/2$). The recurrences of nucleon appear at $J = 1/2, 5/2, 9/2, 13/2, ... $ and their 
masses $m$ are given by Regge formula (\ref{Regge}). 
Quark confinement ensures that such hadronic towers are infinite. 
When proper care is taken of spin dimension the universal slope is equal to
\begin{equation}
\label{alphaprime}
\alpha ' = h/(2 \pi \Delta m^2) \approx 0.378 \times  10^{21}~cm^2/ (g~s)
\end{equation}
(experiment tells us that $\Delta m^2$, which describes the spacing of the masses of Regge recurrences, is approximately equal to proton mass squared).
The slope $\alpha '$ takes care of the string-like character of excited hadrons (the emergence of interquark strings) and permits one to express momenta as {\it proportional} (not inversely proportional) to positions via the dimensional constant
\begin{equation}
\label{kappaR}
\kappa_R = c^2 /\alpha' = 2.37~g/s,  
\end{equation}
which describes string tension and
has the dimension of momentum/position (ie. $(g~cm/s)/cm$ ). \\

It should be clear that the mass scale of ground state hadrons (which is of the order of $10^{-24}~g$), 
and the slope of Regge trajectories (with $\kappa_R$ of the order of a few $g/s$) are two a priori independent parameters. 
After all, with $m_n$ (or $m_{\pi}$) fixed, the relevant
Regge trajectories could still be very steep or very flat (with $\Delta m^2 << m_n^2$ or 
$\Delta m^2 >> m_n^2)$.
Yet the Regge scale of Eq. (\ref{kappaR}) satisfies a relation somewhat
similar to the Eddington-Weinberg relation of Eq. (\ref{N1/3}).
Indeed, we observe that from the four fundamental constants $h$, $G$, $c$, and $\Lambda$ one can form two constants of dimension $g/s$:\\

the `classical' constant
\begin{equation}
\label{kappaC}
\kappa_C = c^3/G = 4.04 \times 10^{38}~g/s,  
\end{equation}

and the `quantum' constant
\begin{equation}
\label{kappaQ}
\kappa_Q = h \Lambda = 0.79 \times 10^{-82}~g/s.
\end{equation}
which differ by the $N$ factor of Eq. (\ref{N10to120}). 
From Eq. (\ref{kappaR}) one
 finds that
\begin{equation}
\label{kappaRkappaC}
\kappa_R = 5.9 \times 10^{-39} \kappa_C,  
\end{equation}
with the proportionality factor not far from $N^{-1/3}=0.4\times 10^{-40}$. Thus, up to a factor of $100$,
one has
\begin{equation}
\label{kappaRvalue}
 \kappa_R \approx N^{-1/3}\kappa_C = c^2\left(\frac{h\Lambda}{3G^2} \right)^{1/3},
\end{equation}
which gives a rough estimate of Regge slope:
\begin{equation}
\label{EW2}
\alpha' \approx \left( \frac{3G^2}{h\Lambda}\right)^{1/3}= 55 \times 10^{21} ~cm^2/(g~s).
\end{equation}
The presence
of a factor close to $N^{-1/3}$ in expression (\ref{kappaRkappaC})
constitutes a large
number coincidence that seems to have not been noticed earlier. The fact that the particular combination of $h$, $G$, and $\Lambda$ given in
Eq. (\ref{EW2}) is of the order of $\alpha'$  may be viewed as a counterpart of the Eddington-Weinberg relation of
 Eq. (\ref{N1/3}). Yet, it concerns not just the nucleon, but the whole Regge tower (trajectory) of hadronic resonances.
The appearance of 
the same factor of $N^{\pm 1/3}$ in connection with hadronic parameters $m_n$ and $\alpha'$ points to their common origin.
 \\

With $m_n$ and $l_n$ being both much smaller than the macroscopic masses and distances
it was contemplated quite early that the spacetime-based
description of nature breaks down at the distance scale of hadronic physics  \cite{Zimmerman} 
\footnote{{Strictly speaking,  the classical notions of space and time are valid on the macroscopic level only, or (as put in \cite{Zimmerman}): `in situations where (...) a dense assembly of clocks and rods may be introduced without significant alteration of the physical situation'. Thus, the application of the classical picture of spacetime to hadron-level physics already constitutes a far-reaching extrapolation. It is achieved by way of merging the classical and quantum aspects of reality in a hybrid classical-quantum approach known under the name of quantum field theory \cite{Finkelstein}.} }.
A decade or more after the introduction of quarks this idea was still on the mind
of several physicists. For example, Penrose viewed his spin-network-induced
twistor conceptions on the emergence of spacetime as appropriate at the hadronic scale 
\cite{PenroseSPinNetworks}.
Today we know that hadrons are made of quarks which are conceived as pointlike objects. 
This discovery is usually taken to mean
that there is no close connection between hadronic physics and the idea of the emergence of spacetime. Yet, one may argue
that such a conclusion is premature as it is based on a misidentification of current (admittedly
very successful) field-theoretical description of strong interactions with physical reality. 
One may believe that the current SM description constitutes an idealization that will yield in time to a deeper description of elementary particles, a description more closely associated with the nature of spacetime \cite{Penrose1968}. The search for physics `beyond the Standard Model' does not have to mean `at smaller distances' or `at larger energies'. It may mean  `beyond the field-theoretical framework', or `beyond the current conception of spacetime'.\\

\section{Internal quantum numbers and phase space}
\label{sec4}

\subsection{Nonrelativistic phase space}
The Standard Model contains several ingredients that are put in by hand.
We do not know why there are three generations of fundamental fermions or why 
each generation is composed of two leptons and two sets of three quarks, all these 
particles being characterized by such internal quantum numbers as weak isospin, hypercharge, and color.
It seems that the first problem that should be addressed
 is the issue of this generation structure.\\

With space viewed as an attribute of matter,
properties of macroscopic 3D space and time should be associated with the
properties of elementary particles.
Indeed, the quantum numbers of spin and parity are connected  
with spatial 3D rotations and reflections, while the existence of particles and antiparticles may be linked to the operation of time reversal. With spatial quantum numbers being associated with the nonrelativistic conception of space and time, it is natural to expect 
that their existence could be deduced via some appropriate nonrelativistic analysis. \footnote{Although the existence of particles and antiparticles is generally
viewed as an implication of the relativistic Dirac approach, it may be inferred from nonrelativistic 
analysis as well \cite{HorzelaKapuscik}.} 
{{Indeed, although inclusion of relativity leads to left and right spinors transforming differently under Lorentzian boosts, it does not affect the very existence and nonrelativistic properties of such concepts as spin or parity.} }
At this point it is interesting to note that the above argument concerning the relative 
unimportance of special relativity 
in the study of some aspects of particle physics seems to go hand in hand with the absence of c 
in the expression for $m_N$, which is essentially the hadronic scale. \\

 Given the existence of the nonrelativistic connection between space and spatial quantum numbers,
it is tempting to expect that the internal quantum numbers could be associated with some 
nonrelativistic extension of the 3D space + time picture. 
{Although relativity has to appear at some later stage of any full discussion, its introduction should not affect the main (`nonrelativistic') conclusions of such an  approach.   Furthermore, we think  that it is too early for the inclusion of special relativity at the level of {\it individual} colored quarks. Indeed, we do not understand some important aspects of intra-hadronic physics already at the nonrelativistic level (for more details see Section \ref{Defense}). Thus, for our purposes, we may restrict attention to a nonrelativistic case.}
Now, the existence of dimensional constant $\kappa_R$
permits expressing positions as proportional to momenta, and allows the introduction of 
additional symmetry between these two sets of coordinates of the 6D phase space.  
\footnote{For example, the existence 
of $\kappa$ permits a replacement  of positions by momenta and vice versa 
(actually ${\bf x} \to {\bf p}$, ${\bf p} \to -{\bf x}$), a 'reciprocity' symmetry originally
introduced by Max Born \cite{Born} in connection with the problem of mass.}
With $\kappa$ having dimension [momentum/position], such a symmetry  
does not have much to do with the quantum connection which involves Planck constant of dimension [momentum $\times$ position].  \\

Such a phase-space-based
description 
of reality seems to constitute a {very} natural {and truly} minimal `extension' 
of the standard 3D description. 
{{Indeed, it may be viewed as (in a sense) `a null extension', for it does not introduce {\it any} additional dimensions of (position) space. It is therefore more minimal than, for example,  the approach of \cite{Baylis} which assumes seven spatial dimensions. \footnote{
 {{ In paper \cite{Baylis} the first three dimensions are the (observed) dimensions of position space, while the remaining four are added to make possible such an enlargement of the underlying geometric algebra that would permit -- with the help of the concept of an algebraic spinor -- the incorporation of all eight fermions of one generation of the Standard Model. } } } } }
In addition, the phase space approach  realizes the philosophical condition of a symmetric treatment
of things and processes (i.e. positions of things and their changes or, in other words, motions)
as advocated by such thinkers as Heraclitus, Leibniz, and - most notably - Whitehead \cite{Whitehead,Zenbook}.
The utmost {parsimony} of this extension and its deep philosophical underpinning makes one hope that the expected additional phase-space-related quantum numbers could be identified with some observed internal quantum numbers. 
{Of course, the above arguments should be backed by other arguments, ideally of observational or experimental nature. Such arguments will be provided later on.} \\

\subsection{Linearization}
The simplest argument that leads from the level of macroscopic classical space 
(of either positions ${\bf x}$ or momenta ${\bf p}$) to the level of spatial quantum numbers is supplied by the Dirac linearization idea.  
According to this idea the 3D rotational invariant of momentum
square ${\bf p}^2$  may be written as a product of two identical factors 
${\bf A}\cdot {\bf p} \equiv \sum_{m} A_m p_m$, 
linear in momentum p:
\begin{equation}
\label{LD}
        {\bf p}^2= ({\bf A}\cdot{\bf p})({\bf A}\cdot{\bf p}), 
\end{equation}
where ${\bf A}$ is some momentum-independent object that behaves like ${\bf p}$ under rotations 
(so that ${\bf A}\cdot{\bf p}$ is a rotational invariant) and satisfies
certain requirements that follow from (\ref{LD}). Specifically, as
terms proportional to $p_m p_n$ are absent on the l.h.s. above, 
the $A_m$'s must satisfy anticommutation rules
\begin{equation}
\label{CR}
         A_mA_n+A_nA_m=2\delta_{mn},     
\end{equation}
which means that $A_m$ cannot be represented by ordinary numbers.
As is well known, the above equation may be satisfied if one takes
                    $A_m=\sigma_m$
where $\sigma_m$ are Pauli ($2 \times 2$) matrices.
Thus, linearization connects the rotational properties of the classical 3D macroscopic space with the quantum concept of spin $S_k=\sigma_k/2$. \\

When an extension from the 3D momentum (or position) space to the 6D phase space is investigated,
it is natural to consider 
\footnote{For simplicity in the following
 we measure ${\bf p}$ and ${\bf x}$ in such units that both $\kappa$ and $h$ are $1$.} 
the phase-space analogue of ${\bf p}^2$, i.e. ${\bf p}^2+{\bf x}^2$ 
and to attempt its linearization \`a la Eq. (\ref{LD}):
\begin{equation}
\label{PScl}
              {\bf p}^2+{\bf x}^2 = ({\bf A}\cdot{\bf p}+{\bf B}\cdot{\bf x})
({\bf A}\cdot{\bf p}+{\bf B}\cdot{\bf x}),  
\end{equation}
where $A_m$ and $B_n$ are the analogues of $A_m$ in Eq. (\ref{LD}). 
Let us now look at some details this idea entails. 
With momentum and position coordinates considered as classical (i.e. commuting) variables, the above equality requires six objects
$A_m$ and $B_n$ to satisfy anticommutation conditions analogous to (\ref{CR}),
which define the Clifford algebra of 6D phase space. 
One finds that these conditions
may be satisfied with $A_m$ and $B_n$ represented by $8 \times 8$ matrices \cite{Zenbook, ZenAPPB1, ZenJPConfSer}.
A convenient representation of these six elements is provided by  tensor products of Pauli matrices:
\begin{eqnarray}
A_m &=& \sigma_m \otimes \sigma_0 \otimes \sigma_1,\nonumber\\
\label{representation}
B_n &= &\sigma_0 \otimes \sigma_n \otimes \sigma_2.
\end{eqnarray}
When momenta and positions are considered as quantum variables, 
then  - due to the nonzero value of commutator $[x_m,p_n]=i\delta_{mn}$ - 
 there appears an additional term on the l.h.s. of (\ref{PScl}), i.e. one gets
\begin{equation}
\label{PSlinearization}
 ({\bf A}\cdot{\bf p}+{\bf B}\cdot{\bf x})
({\bf A}\cdot{\bf p}+{\bf B}\cdot{\bf x}) = {\bf p}^2+{\bf x}^2 + R,   
\end{equation}
with 
\begin{equation}
\label{Rdef}
 R = -\frac{i}{2} \sum_k [A_k,B_k] =\sum_k \sigma_k \otimes \sigma_k \otimes \sigma_3.
\end{equation}
After introducing the element
\begin{equation}
\label{Bdef}
B= iA_1A_2A_3B_1B_2B_3 = \sigma_0 \otimes \sigma_0 \otimes \sigma_3,
\end{equation}
multiplying a rescaled version of Eq. (\ref{PSlinearization}) by $B$: 
\begin{equation}
\label{Qdef}
Q \equiv \frac{1}{6}\left[ ({\bf p}^2+{\bf x}^2)_{vac}+R \right] ~B
\end{equation}
(where $({\bf p}^2+{\bf x}^2)_{vac} =3$ is the 
lowest eigenvalue of ${\bf p}^2+{\bf x}^2$),
and defining matrices
\begin{eqnarray}
\label{I3def}
I_3 & = & \frac{1}{2} B, \\
\label{Ydef}
Y   & = & \frac{1}{3} RB = \frac{1}{3} \sum_k \sigma_k\otimes\sigma_k\otimes\sigma_0,
\end{eqnarray}
one obtains 
\begin{equation}
\label{GMN}
Q = I_3+Y/2.    
\end{equation}
It may be checked that the $8 \times 8$ matrices $I_3$ and $Y$ commute among themselves and
are invariant under ordinary 3D rotations and reflections. Consequently, 
their eigenvalues constitute natural 
candidates for internal quantum numbers. 
With these eigenvalues being 
\begin{eqnarray}
I_3 & \to & \pm 1/2,        \phantom{xxxxxxxxxxxxxxxxxx}      ({\rm for~any}~Y), \nonumber\\  
Y & \to & -1, +1/3, +1/3, +1/3,  \phantom{xxxxx.}      ({\rm for~any}~I_3),
\end{eqnarray}
the eigenvalues of $Q$ are $(0, +2/3, +2/3, +2/3, -1, -1/3, -1/3, -1/3)$,  identical with the charges of eight fundamental fermions composing a single generation of the SM. 
It is therefore natural to identify (\ref{GMN}) with the Gell-Mann-Nishijima formula for electric charge $Q$,
and $I_3$ and $Y$ with weak isospin and hypercharge. \\

\subsection{Rishons or `partial hypercharges'}
The triple appearance of the $Y$ eigenvalue of $+1/3$ 
is naturally associated with the triplicity of the color quantum number.
It is instructive to see how this triplicity emerges. 
From (\ref{PSlinearization}) we observe that
the hypercharge $Y$ is built as a sum of three mutually commuting `partial hypercharges'
\begin{equation}
Y_k= - \frac{i}{6}[A_k,B_k]B=  \frac{1}{3}\sigma_k\otimes\sigma_k\otimes\sigma_0.
\end{equation} 
Calculation shows that the values of $Y$ are constructed from $Y_k$ in the way 
indicated in Table \ref{table1}.
\begin{table}[h]
\caption{Decomposition of hypercharge eigenvalues}
\begin{center}
\begin{tabular}{ccccc}\hline
 $Y_1$     &   $Y_2$   &    $Y_3$    &       $Y$     &       particle \rule{0mm}{6mm}\\
\hline
$-1/3$    &  $+1/3$    &   $+1/3$    &     $+1/3$    &      red quark \rule{0mm}{6mm}\\
$+1/3$    &  $-1/3$    &   $+1/3$    &     $+1/3$    &     blue quark \rule{0mm}{6mm}\\
$+1/3$    &  $+1/3$    &   $-1/3$    &     $+1/3$    &    green quark\rule{0mm}{6mm}\\
$-1/3$    &  $-1/3$    &   $-1/3$    &     $-1$      &     lepton \rule{0mm}{6mm}\\
\hline
\end{tabular}
\end{center}
\label{table1}
\end{table}
The rightmost column in Table \ref{table1} provides the correspondence between the structure of $Y$ and the quartet composed of a lepton and three colored quarks. The triplicity of color is attributed here to three different orderings 
in which $Y=+1/3$ may be built out of $-1/3$, $+1/3$, and $+1/3$. \\

It appears \cite{ZenJPConfSer} that the pattern exhibited in Table \ref{table1} is in one-to-one correspondence with the way in which
the charges of leptons and quarks are built out of the charges of their alleged subparticles, the so-called `rishons' of the Harari-Shupe (HS) model \cite{HS}. 
In that model the eight fermions of a single SM generation
are conceived as ordered triplets of two `truly fundamental' spin-1/2 subparticles (rishons) $T$ and $V$ of charges $+1/3$ and $0$ respectively. For example, the red $u$ quark is identified with the triplet $VTT$, 
while the neutrino $\nu$ - with $VVV$. With only two types of rishons
the HS scheme is very economic in the number of fundamental particles.
Yet, it exhibits {many} shortcomings in other places. 
In particular, it predicts the existence of unobserved particles 
(e.g. spin-3/2 partners of leptons and quarks), its concept of color is not connected with the $SU(3)$ color group of the SM, it violates the condition of fermion antisymmetrization at the rishon level, etc., etc. 
It turns out that in the phase-space framework these shortcomings of the HS rishon model do not appear.  
The basic reason is that in that approach the partial hypercharges (the counterparts of HS rishons) 
constitute algebraic components of the hypercharge operator only. 
They do not reside on any subparticles.  Lepton and quarks of a single SM generation are connected by phase-space-induced symmetry, 
but its explanation in terms of subparticles is neither needed nor possible.  
For a more detailed presentation of the relevant arguments see \cite{ZenGlimpse}. 
By Ockham's razor, the phase-space explanation of the observed pattern of fundamental fermions is vastly superior to that provided by the original rishon model. \\

The shortcomings of the  HS model result from the intuitive wish to divide
matter again and again. Yet, the divisibility of matter must come to an end. As noted by Heisenberg \cite{HeisenbergPhysicsToday}, this comes about as a change in the meaning of the word 'to divide' so that after several steps down the ladder of compositeness the concept of further subdivision loses most of its original meaning. 
One of the lowest and most important steps of this ladder seems to occur  
when the concept of macroscopic separability is lost, i.e. during the transition
from the hadronic to the quark level. As the phase-space scheme suggests,
going further down (to the supposed rishon level) seems to require a strictly algebraic understanding of the compositeness of matter (i.e. the construction of hypercharge from `partial hypercharges' that do not reside on subparticles) and, consequently, the {\it inapplicability of the ordinary conception of matter below the lepton/quark level}. 
 The conceptual superiority of the phase-space scheme over that provided by the original HS model strongly suggests that
the ordinary conception of matter starts at the lepton/quark scale, not below it.
With space understood as an attribute of matter, it seems therefore
that it is the transition from the partial hypercharge ('rishon') level
 through the lepton/quark level 
and on to the hadronic level (i.e. the emergence of matter) that should be relevant for the understanding of 
the emergence of space. The Planck scales seem quite irrelevant here.\\

\section{Generalization of mass, emergence of hadrons} 
\label{sec5}

\subsection{Problem of mass}
The problem of mass is a thorny one. 
For a long time the main controversy seemed to have been between the view that mass 
is an intrinsic property of a given particle and the opinion that it results from 
the interaction of that particle with other objects. According to 
Wigner \cite{Wigner},
from the group-theoretical standpoint the observed particles may be labelled by two
space-related properties, namely spin and mass, associated with 
rotations and translations respectively. 
This seems to suggest similar physical origins of mass and spin. 
Yet, the current view treats spin and mass somewhat asymmetrically: it holds
that while spin is an intrinsic property of a particle, it is the interaction with the Higgs field that generates the masses of
originally massless particles. \\

Still, the Higgs mechanism (as it stands now) does not provide us 
with a substantial improvement in our understanding of the problem of mass.  
As noted in \cite{Hanson}, it {\it `merely replaces one set of unknown parameters (particle masses) with an equally unknown set of parameters (coupling constants to the Higgs field(s)), 
so nothing is gained in the fundamental understanding of masses.'} 
The problem of mass is further exacerbated by the existence of mixing between different generations.
Thus, we need a principle deeper than the Higgs mechanism.  
It should provide a single rationale behind the existence of the whole 
variety of fundamental particles, explain the observed pattern of their masses 
and mixings, and reduce the number of 
free SM parameters. For example, 
Hansson \cite{Hanson} argues that the relative 
values of neutrino, electron, and quark masses are correlated with the strength of fundamental 
interactions in which these particle participate. 
Given our ignorance as far as origin of mass is concerned it is obvious that one should welcome any additional light that could be shed
on this problem.
\\
 
\subsection{Phase space and mass}
The phase-space scheme provides an attractive explanation of the origin of several internal quantum numbers. It turns out that it has other interesting implications as well. They bear on the idea of space as a property of matter. Generalization of the concept of mass is one of them. It is this generalization that (as we believe) may help us in future to uncover how {classical macroscopic} space emerges from {(and is constructed as a limiting property of)} the underlying {material} quantum layer. 
It was observed by Max Born \cite{Born} that the ordinary concept of mass distinguishes between momentum and position variables. Indeed, mass of individual free physical bodies always enters into dispersion formulas in association with momentum (and not with position), be it in a nonrelativistic or relativistic 
expressions for particle energies i.e.  
\begin{equation}
\label{standard}
E={\bf p}^2/2m ~~{\rm or}~~ E^2={\bf p}^2+m^2.
\end{equation}
Born's reciprocity symmetry between ${\bf p}$ and ${\bf x}$ violates this association of mass with momentum.  

The introduction of $\kappa$ permits a completely parallel treatment of momentum and position variables and suggests the introduction of phase-space invariant ${\bf p}^2+{\bf x}^2$.  This invariant admits the consideration 
of transformations that replace momenta with positions in various ways, not only via Born's reciprocity transformation. Specifically, consider the following 
rotations in phase space:
\begin{eqnarray}
p'_1 &=& p_1 \cos \phi + x_3 \sin \phi, \nonumber \\
x'_3 &=& x_3 \cos \phi - p_1 \sin \phi, \nonumber \\
p'_2 &=& p_2, \nonumber \\
x'_2 &=& x_2, \nonumber\\
p'_3 &=& p_3 \cos \phi - x_1 \sin \phi, \nonumber\\
\label{rotPhSp}
x'_1 &=& x_1 \cos \phi + p_3 \sin \phi.      
\end{eqnarray}
Invariance of ${\bf A}\cdot{\bf p}+{\bf B}\cdot{\bf x}$ requires that $A_m$ and $B_n$ transform as above 
with $p_m \to A_m$ and
$x_n \to B_n$, and with similar replacements for primed objects.
For $\phi = \pi/2$ one then immediately finds that 
\begin{eqnarray}
A'_1 = B_3,  \phantom{xxx}     &     A'_2 = A_2,  \phantom{xxx}   &     A'_3 = -B_1, \nonumber \\
\label{AtoBandviceversa}
B'_3 = -A_1, \phantom{xxx}     &     B'_2 = B_2,   \phantom{xxx}  &     B'_1 = A_3,
\end{eqnarray}
and, consequently, 
\begin{eqnarray}
B'  &=& B, \nonumber \\
Y'_1 &=& - Y_3, \nonumber \\
Y'_2 &=& Y_2, \nonumber \\
\label{BYktransf}
Y'_3 &=& - Y_1,
\end{eqnarray}
i.e. nothing happens to the values of $I_3=B/2$ and $Y_2$,
while the values of $Y_1$ and $Y_3$ are interchanged (with additional negative signs).
This leads to Table \ref{table1} being transformed (row by row) into Table \ref{table2}.
\begin{table}[h]
\caption{Structure of $Y$ after rotation in phase space}
\begin{center}
\begin{tabular}{ccccc}\hline
 $Y'_1$     &   $Y'_2$   &    $Y'_3$    &       $Y'$     &       particle \rule{0mm}{6mm}\\
\hline
$-1/3$    &  $+1/3$    &   $+1/3$    &     $+1/3$    &      red quark \rule{0mm}{6mm}\\
$-1/3$    &  $-1/3$    &   $-1/3$    &     $-1$    &     lepton \rule{0mm}{6mm}\\
$+1/3$    &  $+1/3$    &   $-1/3$    &     $+1/3$    &    green quark\rule{0mm}{6mm}\\
$+1/3$    &  $-1/3$    &   $+1/3$    &     $+1/3$      &     blue quark \rule{0mm}{6mm}\\
\hline
\end{tabular}
\end{center}
\label{table2}
\end{table}
We see that the lepton and the blue quark are interchanged, while
the red and green quarks are unaffected. One may say that a quark of a given color 
is a lepton appropriately
rotated in phase space. \\

Consider now the effect that this transformation has on 
the connection between phase space variables and the classical concept of mass.
Let us focus on the standard dispersion relation for a macroscopic body, that is
naturally extrapolated to free microscopic objects, such as a lepton. 
Then, according to Eq. (\ref{rotPhSp}), in the  dispersion formula appropriate 
for the blue quark,  the ordinary three-momentum $(p_1,p_2,p_3)$ is replaced
 by a `mixed' triplet of phase space variables, i.e. by 
a new canonical momentum
\begin{equation}
\label{mixed}
(p'_1, p'_2, p'_3) = (x_3, p_2, -x_1).
\end{equation}
Thus, a free blue quark would have to satisfy a symmetry-related phase-space 
counterpart of Eq.(\ref{standard}) i.e. (for an originally relativistic relation)
\begin{equation}
\label{blue}
E^2= x_3^2+p_2^2+x_1^2+m^2.  
\end{equation}
By analogy, similar mixed triplets and dispersion formulas (with cyclic relabelling 
$2 \to 3 \to 1 \to 2$) appear for the red and green quarks. \\

It may be objected that these implications of the phase-space scheme 
are totally unacceptable as they violate rotational,
translational, and relativistic invariances of the surrounding macroscopic world. 
We think, however, that the appearance of such violations 
is not a vice but one of the greatest assets of the proposed scheme.  
In our view, such would-be violations lie at the origin of quark confinement 
(or - more conservatively speaking - of a novel perspective on it). 
 In other words, we believe that it is due to the would-be violation of rotational and 
translational invariances by the quark dispersion formula that a single quark cannot be observed 
in the familiar classical 3D world in which these invariances obviously have to hold. 
Thus, unobservability of individual quarks is {\it predicted} by the scheme. \\

The proposed explanation of quark confinement seems to be in a manifest contradiction with the generally accepted standard picture of quarks and their confinement via gluon-mediated long-range 
QCD interactions. Therefore, a first reaction to the phase-space picture could be  to discard it right away. 
Yet, before one makes such a hasty decision, one should discuss possible ways
of resolving the conflict.
In particular, one should first address such issues as
the possible connection between phase-space ideas and 
the ordinary description of hadron substructure in terms of quantum quark fields, 
and the disagreement with the widely embraced standard view on quark masses. 
Then, one may turn to the discussion of a seemingly unacceptable description of quark confinement.\\
 
\subsection{Defense of phase space picture}
\label{Defense}
First, let us stress that in our phase-space considerations 
we are concerned with the dispersion relation of a free quark  
(i.e. in the precise association of the concept of quark mass with
phase-space variables), not with other quark properties such as e.g. quark spin, chiral properties, etc.
which are assumed to be fully relevant for the field-theoretical description of {the interaction of quarks with external probes such as photon or weak bosons}. 
The non-standard dispersion relation that an individual quark is supposed to satisfy is viewed as a classical constraint that should be introduced into the quantum field-theoretical framework 
from the `outside' (i.e. from the classical level, just as it is done in the case of the standard dispersion formula).
The fact that {in our approach the} individual quarks are expected to subscribe to a non-standard dispersion relation does {\it not} mean 
that in the quantum field-theoretical approach (and from the point of view of the 3D space + time background {and color-blind external probes}) they cannot be described with the help of a pair of
ordinary (left and right) spinorial fields (the bispinor $q(x)$), as it is standardly done in quark/hadron phenomenology and in QCD. A large part of standard 
phenomenological and field-theoretical description of hadrons in terms of quark substructure is thereby automatically accepted.
What should be skipped are those parts of that picture
 which involve the classically-motivated on-mass-shell 
constraints such as e.g. the Dirac equation ($\slashed{p}-m)q(x)=0$. \cite{ZenJPConfSer}
 This is highly welcome as the use of ordinary
on-mass-shell formulas for quarks is unacceptable on conceptual grounds: 
in the standard field-theoretical approach the ordinary on-mass-shell condition ($p^2=m^2$) corresponds 
to {\it spatial infinity} which obviously cannot be reached by confined quarks.
For similar reasons I find it hard to accept without reservations those quark-level calculations which involve standard quark propagators $1/(\slashed{p}-m)$: 
after all, the very idea of confinement forbids the existence of quark poles at $p^2=m^2$. In my view, such calculations have to be treated as approximations 
that may lead to a variety of artefacts and - for these reasons - cannot be really trusted.
 And indeed, not only are the standard quark propagators   
in conceptual conflict with the very unobservability of free quarks, but their 
carefree use does lead to incorrect predictions also in other, less obvious places (see eg. \cite{ZenWRHD2006}). \\

Second, when confronted with the above claim that the standard concept of mass is not appropriate for a quark,
most physicists would presumably argue that our understanding of quark masses cannot be that bad: they would be inclined to belief in the soundness of standard quark mass extraction procedures. 
After all, respectable scientific literature accepts standard (Dirac) character of quark masses and lists quite precise values for these parameters. 
The point is, however, that - with free quarks being unobservable in asymptotic states - any extraction of quark
mass from experimental data must depend heavily on theory, i.e. on the way quark `mass' is defined and built into the relevant theoretical scheme. Obviously, there is no doubt that a quark may be assigned some 
effective mass parameter that can be extracted from hadronic level data. 
Yet,  what is actually being extracted via the relevant theoretical procedures?  
Are the extracted quark masses the Dirac masses? Or are they mass-like parameters of a more general nature? Do we know the relevant theory sufficiently well for such extractions to be fully reliable?
It turns out that the original,  fifty years old prescription for the extraction of the so-called 
`current' quark masses (via the `current algebra' of hadronic currents) does not actually treat quarks 
as Dirac particles `moving' within hadrons. 
\footnote{For a brief discussion of how the standard values of quark masses are extracted from 
experiment via `current algebra' see eg. \cite{ZenJPConfSer}.} 
For our purposes it is sufficient to say that in these extractions one
uses {\it global} chiral properties of effective quark mass terms as well as spacetime concepts defined 
at the {\it hadronic} level. Some quark-level symmetries do enter into the game 
but no assumption is made concerning the existence of ordinary background 
space within hadrons. Moreover, in such extractions the on-mass-shell Dirac condition
$(\slashed{p}-m)q(x)=0$ does not have to be used (although it sometimes is), thus making it possible to avoid the pending conceptual 
conflict between its use and quark confinement. \\

 More modern quark mass extractions are based on lattice QCD calculations 
in which, although the quarks are pictured on the background of classical spacetime, standard quark propagators do not appear. Thus, just as in the case of current algebra, 
there emerges no conceptual conflict with 
the unobservability of individual quarks in asymptotic states. The mass values extracted via such more sophisticated calculations are in good agreement with the old current algebra estimates. 
This seems to suggest that one may go beyond the simple current algebra picture
and extend the standard spacetime concepts right into the hadronic `interior'. Accordingly, the concept of quark mass may be regarded as quite standard,
with its Dirac nature being masked by QCD confinement effects.
Yet, we shall point out below that this extension seems to go too far. \\
 
Before we turn to the issue of the essential difference in the pictures of confinement, let us comment on some interesting aspects of the phase-space-induced view on the notion of mass
that are missing in the SM approach. Namely, a  study of all 64 elements of the Clifford algebra of phase space shows \cite{ZenClifford,Zenbook} that there is only one element of this algebra (up to Born's reciprocity transformation) that may be associated with the concept of lepton mass. Its rotations in phase space / Clifford algebra (using 
Eq. (\ref{AtoBandviceversa}) etc.) lead to three Clifford algebra elements that have to be associated with the 
concepts of mass for tri-colored quarks.  It turns out that these elements are diagonal elements of a rank-2 tensor that is symmetric in 3D indices \footnote{While Clifford algebras in general lead to antisymmetric tensors only, the phase-space-induced doubling of the 3D structure (the parallel treatment of position and momentum spaces) leads to elements that - from the point of view of our 3D world - are symmetric in 3D indices.}, possibly hinting at the connection of quarks 
with the appearance of metric and thus with the idea of space emergence.
Furthermore, the rotations in phase space involve string tension $\kappa_R$.
This seems to indicate that the problem of mass does not really break up into two 
independent problems: the problem of the mass of fundamental fermions 
(Clifford algebra elements for lepton and quark masses) 
and the problem of the strength of interquark confining interactions (string tension,
$\kappa_R$), 
as it is often thought nowadays.
Instead, these two problems seem to constitute two closely related parts of a single puzzle. 
\\

Finally, we come to the third point - the issue of confinement and the difference between the
phase-space view and the QCD picture. The conceptual simplicity {and parsimony} of
the phase space approach strongly suggests that this approach contains several important elements of truth. If so, we should not reject it immediately but rather find a way to
marry it with the QCD picture.
This may be attempted if one accepts the view that
both the phase-space ideas
and the QCD picture  are idealizations 
that may deviate from reality in various places. \footnote{
After all, all our theories are abstract descriptions, models built 
with the goal of representing various features of nature and applicable to its limited 
regions only \cite{Heisenb1958}.
They must not be treated as the underlying and complete truths 
that are valid everywhere. }
Thus, the phase-space and QCD pictures should be viewed as
two different perspectives on the nature of quark confinement. As such they may benefit from each other. It is therefore
gratifying that the two pictures exhibit important similarities. 
For example, the phase space approach provides the justification for the appearance of the SU(3) color symmetry group of QCD.
Furthermore, both the phase-space picture and QCD involve the description of confinement
in terms of strings, even though these strings are pictured differently. \\
 
{Addressing the question of how to marry the phase-space view with the QCD picture requires going beyond the intended scope of the present paper. Nonetheless, we should mention here an idea that was put forward in \cite{Mulders} and could be relevant to this question. The proposal of \cite{Mulders} bears some resemblance to the phase-space scheme: it links the number of colors to the number of space dimensions, constructs leptons and quarks from one-dimensional structures somewhat akin to our interpretation of Harari-Shupe rishons, and does away right from the beginning with the confinement problem. At the same time it involves left and right objects, dynamical gluons, and more. 
 An intriguing question is if such ideas can be adapted to our scheme, marrying it with QCD and bringing in a way to incorporate special relativity.} \\

Now, while  the phase-space picture (as it stands now) misses important elements of QCD (eg. gluons), QCD also seems to be deficient, although from a different angle.  
In particular, there are strong phenomenological indications from
baryon spectroscopy that something important is missing in the original quark model/ QCD picture.
It appears that the standard quark model and lattice QCD both predict the existence of many more
excited baryonic states than experimentally observed \cite{Capstick}. More specifically, 
it seems that in excited baryons one internal spatial degree of freedom is frozen 
(this provides a direct suggestion that there is  a close
connection of quark-to-hadron transition, {the mechanism of quark confinement,} and 
hadronic scale with the idea of space emergence). 
On this issue Capstick and Roberts write \cite{Capstick}: {\it `If no new baryons are found, 
both QCD and the quark model will have made incorrect predictions, 
and it would be necessary to correct the misconceptions that led to these predictions.
Current understanding of QCD would have to be modified and the dynamics within the quark
model would have to be changed'}. 
We conclude that it is not sufficient to calculate  
the masses of the ground-state mesons and baryons and to claim - 
on the basis of their approximate agreement with experimental values - that the relevant theoretical scheme is correct.
One cannot view such successes of lattice QCD 
as a sufficient argument favoring the existence of standard spacetime background in hadronic `interior'. 
The real challenge for the lattice QCD is the description of baryonic excited states.
It is only after this is achieved that one can accept the applicability of unmodified QCD to
the description of quark confinement.
As the situation stands now, the apparent freezing of one spatial degree of freedom in excited baryons strongly suggests that the nature of hadronic `interior' may be different from 
naive extrapolations. 
We repeat: the fact that one can describe quarks as quantum fields on the background of classical
continuous spacetime and that many interesting conceptions (including various ideas on quark/hadron transition) were developed within such a picture does not mean that this background, assigned to the interior of hadrons by an extrapolation from the macroscopic domain, is {\it fully} adequate for a deeper description of quarks and hadronic structure.
\\

Now, it is interesting to note that on the issue of hadronic `interior'
 the phase-space picture seems to differ from standard quark approaches.
The implication that masses of individual quarks should enter into rotationally 
and translationally non-invariant dispersion relations means that quarks cannot exist as individual free objects. It does not mean, however, that they cannot exist as inseparable components of multi-quark
conglomerates provided these conglomerates satisfy all the invariances that ordinary
 objects are supposed to satisfy.  
The phase-space approach seems to admit the
emergence of such conglomerates. Indeed, although in this approach a precise prescription for the construction
of hadronic states is missing, an interesting argument that leads to the emergence of mesonic and baryonic structures may be given. This argument is based on an extension
of the principle of additivity of physical momenta of ordinary macroscopically separable objects.
According to this principle, the total momentum of a composite system of ordinary objects is given by the sum of the momenta of its
components (${\bf p}_{tot} = \sum_k {\bf p}^{(k)}$).
In the phase space picture one may expect this trivial principle to be generalized
  to the principle of additivity of canonical momenta. This leads in particular to 
the addition of canonical momenta of three quarks of different colors, i.e. of
\begin{eqnarray}
(p^B_1, p^B_2, p^B_3)& = &(-y_1, p_2, x_3), \nonumber\\
(p^R_1, p^R_2, p^R_3)& = &(x_1, -y_2, p_3), \nonumber\\ 
(p^G_1, p^G_2, p^G_3)& = &(p_1, x_2, -y_3),
\end{eqnarray}
and suggests the appearance of a translationally invariant expression
\begin{equation}
\label{Deltastring}
(p_1, p_2, p_3, x_1-y_1, x_2-y_2, x_3-y_3),
\end{equation}
which can be made rotationally covariant if the three quarks conspire so that
${\bf p}$ and ${\bf x} - {\bf y}$ are actually ordinary vectors. 
Thus, using the additivity principle one may construct expressions 
that satisfy the condition of proper rotational and translational behavior at the composite level,
and are therefore appropriate for the description of ordinary objects.
Accepting this generalization of the additivity principle leads to the emergence
of expressions appropriate for the description of mesons and baryons, in a way somewhat similar to their 
standard group-theoretical description.   
Further discussion of this idea may be found in \cite{Zenbook,ZenJPConfSer,ZenClifford}.
Here, we would just like to point out that the additivity prescription treats the two a priori possible internal baryonic spatial degrees of freedom in an asymmetrical manner 
(as only one vector of internal displacement is present in Eq. (\ref{Deltastring})),  
which may have something to do with the apparent freezing of one internal degree of freedom in excited baryons. Thus, the phase space picture - if developed 
for the description of hadrons
in an as yet unknown way - may provide hints on how the QCD picture should be modified.
Obviously, with QCD requiring a prior acceptance of the existence of an underlying
background spacetime, introduction of gluons into the phase-space approach should be postponed until 
a working idea on the emergence of spacetime is proposed.

\section{Conclusions}
\label{sec6}

In conclusion, I think that important information on the idea of space emergence could be extracted from the hadronic realm. There is a couple of arguments that support this point of view. \\

First, with space viewed as an attribute of matter, it seems that it is the discrete spectrum of masses of elementary particles that should define 
discrete properties of space in the microcosm. Thus, one should start not from discrete (or quantized) space but from discrete (or quantized) matter. Hadrons seem particularly relevant here as their spectrum comprises objects of all spins. \\

Second, the standard dimensional argument that singles out Planck mass scale (and the related distance scale) as relevant for space emergence has a natural counterpart that   singles out the hadronic mass scale instead of Planck scale. Indeed, if one accepts four fundamental constants: $h$, $G$, $c$, and $\Lambda$, then the Planck mass scale $m_P$ is obtained if $\Lambda$ is not used, while the hadronic mass scale $m_N$ is obtained if $c$ is not used. 
As the hadronic mass scale $m_N$ is much farther from the classical realm than the Planck mass scale (which is essentially of classical size) it seems that it is $m_N$ 
(and not $m_P$) that should be relevant for the consideration
of spacetime emergence.
Moreover, in addition to $m_N$ there is another a priori independent scale 
(the Regge tension $\kappa_R$)
that characterizes
the hadronic spectrum and is far from the classical realm ($\kappa_R << \kappa_C $). \\ 

Third, the existence of hadronic spectrum parameter $\kappa_R$ supports a phase-space picture that
provides a justification 
for the emergence of internal quantum numbers
of weak isospin, hypercharge and color.  
{As this picture is essentially unique in its parsimony,} it
is very attractive. It suggests an extension of the concept of mass (and a more general view on matter
that involves a symmetric treatment of things and processes) and 
predicts the unobservability of individual quarks.
It also seems to involve an asymmetric treatment of the two intra-baryonic spatial degrees of freedom, which are expected in simple extrapolations of the macroscopic conception of space to the hadronic `interior'. Thus, it questions the idea of ordinary divisibility of space when going from the hadronic to the quark level, and suggests that standard ideas on intra-baryonic space constitute an approximation to reality only. 
Such a view on hadronic `interior' is corroborated by
phenomenological analyses of the spectrum of excited baryons which indicate
that one internal spatial degree of freedom is frozen.\\

Fourth, 
extending the idea of the divisibility of matter below the lepton/quark level,
as it is assumed in the rishon model, leads to many shortcomings which are automatically absent if 
one adopts a strictly algebraic interpretation of rishons as obtained in the phase-space picture. \\
 
To summarize, I think that the emergence of space may and should be viewed as a byproduct of
the transition of matter from the (algebraic) rishon level via the (particle/quasi-particle)
lepton/quark level and on to the (particle) lepton/hadron level. 
The currently dominant view of continuous space emerging at the Planck distance scale
seems to be in conflict with the philosophical position accepting logical priority of matter over space when this position is combined with the analysis of the salient properties of (quantized) matter.\\

\vfill

\vfill


\begin{thebibliography}{99}
\bibitem{Heisenberg1979} Heisenberg W., Ideas of the natural philosophy of ancient times in modern physics, in
{\it Philosophical Problems of Quantum Physics} (OxBow Press, Woodbridge, Connecticut, 1979).
\bibitem{Wuthrich} See eg. W\"uthrich C., To quantize or not to quantize: Fact and folklore in quantum gravity, Phil. Sci. 72 (2005) 777-788.
\bibitem{Amelino} Amelino-Camelia G., Quantum gravity phenomenology, physics/0311037; Baez J. C., Higher-dimensional algebra and Planck-scale
physics. In Callender C. and Huggett N. (eds), {\it Physics meets philosophy at the Planck scale},
(Cambridge Univ. Press, 2001), p.177-195.
\bibitem{Meschini} Meschini D., Planck-scale physics: facts and beliefs, arXiv: gr-qc/0601097, Found. Sci. 12 (2007) 277-294.
\bibitem{Finkelstein}   Finkelstein D. R., Space-time code, Phys. Rev. 184 (1969) 1261-1271; Space-time code. 2, Phys. Rev. D5 (1972) 320-328.
\bibitem{Bojowald} Bojowald M., Quantum gravity and cosmological observations, arXiv: gr-qc/0701142, AIP Conf. Proc. 917 (2007) 130-137.
\bibitem{Wesson} Wesson P. S., Is mass quantized?, arXiv: gr-qc/0309100,  Mod. Phys. Lett. A19 (2004) 1995-2000.
\bibitem{Thai} Burikham P., Dhanawittayapol R., and Wuthicharn T., A new mass scale, implications on black hole evaporation and holography
, arXiv: 1605.05866, 
Int. J. Mod. Phys. A31 (2016) no.16, 1650089.
\bibitem{EW} Eddington A. S., {\it Fundamental Theory} (Cambridge University Press, Cambridge, England, 1946); Weinberg S., {\it Gravitation and Cosmology} (Wiley, New York, 1972).
\bibitem{Funkhouser} Funkhouser S., A new large-number coincidence and a scaling law for the cosmological constant, arXiv: physics/0611115, Proc. Roy. Soc. A464 (2008) 1345-1353.
\bibitem{Zimmerman} Zimmerman E. J., The macroscopic nature of space-time, Am. J. Phys. 30 (1962) 97-105.
\bibitem{PenroseSPinNetworks} Penrose R., Twistors and particles: an outline, in {\it Proc. of the Conference on 
Quantum Theory and the Structures of Time and Space} (Feldafing, July 1974), p. 129-145.
\bibitem{Penrose1968} Penrose R., Structure of spacetime, in DeWitt C. M. and 
Wheeler J. A. (eds), {\it Batelle Rencontres} (New York, 1968), p. 121-235.
\bibitem{HorzelaKapuscik} Horzela A. and Kapu\'scik E., 	
Galilean covariant Dirac equation, Electromagnetic Phenomena V.3 (2003) 63-69.
\bibitem{Born} Born M.,  Reciprocity theory of elementary particles, Rev. Mod. Phys. 21 (1949) 463-473.
\bibitem{Baylis} Trayling G. and Baylis W., A geometric basis for the standard-model gauge group, J. Phys. A34 (2001) 3309-3324.
\bibitem{Whitehead} Irvine A. D., Alfred North Whitehead, in Edward N. Zalta (ed.) {\it The Stanford Encyclopedia of Philosophy} (Winter 2010), \texttt{http://plato.stanford.edu/archives/win2010/entries/whitehead/}; Kraus E., {\it The Metaphysics of Experience: a Companion to Whitehead's Process and Reality} (Fordham University Press, New York, 1979).
\bibitem{Zenbook} \.Zenczykowski P., {\it Elementary particles and emergent phase space}, (World Scientific, Singapore, 2014).
\bibitem{ZenAPPB1} \.Zenczykowski P., Space, phase space and quantum numbers of elementary particles, Acta Phys. Pol. B38 (2007) 2053-2076; The Harari-Shupe preon model and nonrelativistic quantum phase space, Phys. Lett. B 660 (2008) 567-572.
\bibitem{ZenJPConfSer} \.Zenczykowski P., Elementary particles, the concept of mass, and emergent spacetime, J. Phys. Conf. Ser. 626 (2015) no. 1, 012022.
\bibitem{HS} Harari H., A schematic model of quarks and leptons, Phys.Lett.B86 (1979) 83-86; Shupe M., A composite model of leptons and quarks, Phys.Lett.B86 (1979) 87-92.
\bibitem{ZenGlimpse} \.Zenczykowski P., The Harari-Shupe observation without preons - a glimpse of physics to come?, arXiv: 1503.07773, Acta Phys. Pol. B47 (2016) 1011-1032.
\bibitem{HeisenbergPhysicsToday} Heisenberg W., The nature of elementary particles, Phys. Today 29 (1976) 32-39.
\bibitem{Wigner} Wigner E. P., On unitary representations of the inhomogeneous Lorentz group, Ann. Math. 40 (1939) 149-204.
\bibitem{Hanson} Hansson J., Physical origin of elementary particle masses, arXiv: 1402.7033, Electron. J. Theor. Phys. 11 (2014) 87-100; On the origin of elementary particle masses, arXiv: 1211.3136, 
    Prog. Phys. 10 (2014) 71-73.
\bibitem{ZenWRHD2006} \.Zenczykowski P., Joint description of weak radiative and nonleptonic hyperon decays in broken SU(3), Phys. Rev. D73 (2006) 076005.
\bibitem{ZenClifford} \.Zenczykowski P., Clifford algebra of nonrelativistic phase space and the concept of mass, J. Phys. A42 (2009) 045204; From Clifford algebra of nonrelativistic phase space to quarks and leptons of the Standard Model, Adv. Appl. Clifford Algebras 27 (2017) 333-344.
\bibitem{Heisenb1958} Heisenberg W., {\it Physics and Philosophy: The Revolution in Modern 
Science} (Harper and Row, New York, 1958) p. 200.
\bibitem{Mulders} Mulders P., The 3D entangled structure of the proton; transverse degrees of freedom in QCD, momenta, spins and more, arXiv:1801.03664, Few Body Syst. 59 (2018) no.2, 10; The 3D structure of QCD and the roots of the Standard Model
, EPJ Web Conf. 112 (2016) 01014.
\bibitem{Capstick}  Capstick S., Roberts W., Quark models of baryon masses and decays,
Prog. Theor. Part. Nucl. Phys. 45 (2000) S241-S331.
\end{thebibliography}
\end{document}